\documentclass[letterpaper,10 pt,conference]{ieeeconf}
\IEEEoverridecommandlockouts  
\usepackage[utf8]{inputenc}
\usepackage[left=1in,top=1in,right=1in,bottom=1in]{geometry}
\usepackage{graphicx}
\usepackage{epsfig}
\usepackage{placeins}
\usepackage{enumerate}
\usepackage{soul}
\usepackage{array}
\usepackage{amsmath}
\usepackage{amssymb}
\usepackage{subcaption}
\usepackage{float}
\usepackage{gensymb}
\usepackage{eqnarray,amsmath}
\makeatletter
\let\NAT@parse\undefined
\makeatother
\usepackage{cite}
\usepackage{breqn}
\usepackage[colorlinks=true,linkcolor=black,anchorcolor=black,citecolor=black,filecolor=black,menucolor=black,runcolor=black,urlcolor=black]{hyperref}
\usepackage{booktabs}
\usepackage{lipsum,amsmath,multicol}
\def\barr{\begin{array}}
\def\earr{\end{array}}
\def\berr{\begin{eqnarray}}
\def\err{\end{eqnarray}}
\def\berrno{\begin{eqnarray*}}
\def\errno{\end{eqnarray*}}
\def\be{\begin{equation}}
\def\ee{\end{equation}}

\usepackage{enumerate}
\usepackage[ruled, linesnumbered, vlined]{algorithm2e}
\SetKwRepeat{Do}{do}{while}%
\usepackage{placeins}
\usepackage{float}
\usepackage{dblfloatfix}
\usepackage{xcolor}
\usepackage{lipsum}
\usepackage{tikz,lipsum,lmodern}
\usepackage[most]{tcolorbox}
\usepackage{array}

\usepackage{placeins}
\usepackage{multirow}
\usepackage{soul}
\usepackage{amsmath}
\usepackage{mathtools, cuted}

\newcolumntype{L}[1]{>{\raggedright\let\newline\\\arraybackslash\hspace{0pt}}m{#1}}
\newcolumntype{C}[1]{>{\centering\let\newline\\\arraybackslash\hspace{0pt}}m{#1}}
\newcolumntype{R}[1]{>{\raggedleft\let\newline\\\arraybackslash\hspace{0pt}}m{#1}}

\hypersetup{%
    pdfborder = {0 0 0}
}
\title{A Comparative Analysis of the Cobb-Douglas Habitability Score (CDHS) with the Earth Similarity Index (ESI)
} 
\author{Surbhi Agrawal$^{1}$, Suryoday Basak$^{1}$, Snehanshu Saha$^{1}$, Kakoli Bora$^{2}$ and Jayant Murthy$^{3}$ 
\thanks{$^{1}$ The Department  of  Computer Science,
        PES University,Bangalore
        {\tt\small snehanshusaha@pes.edu}}%
\thanks{$^{2}$The Department of Information Science,
        PESIT-BSC,Bangalore}     
\thanks{$^{2}$The Department of Information Science,
        PESIT-BSC,Bangalore}
\thanks{$^{2}$Indian Institute for Astrophysics,Bangalore}}      
        
\begin{document}
\maketitle
 %
%
%

\section{Introduction and Methods}

A sequence of recent explorations by Saha et al. \cite{ASCOM18} expanding previous work by Bora et al.\cite{BoraCDHS2016} on using Machine Learning algorithm to construct and test planetary habitability functions with exoplanet data raises important questions. The 2018 paper analyzed the elasticity of their Cobb-Douglas Habitability Score (CDHS) and compared its performance with other machine learning algorithms. They demonstrated the robustness of their methods to identify potentially habitable planets from exoplanet dataset. Given our little knowledge on exoplanets and habitability, these results and methods provide one important step toward automatically identifying objects of interest from large datasets by future ground and space observatories. The paper provides a logical evolution from their previous work. Additionally, model and the methods yielding a new metric for "Earth-similarity" paves the way for comparison with its predecessor, ESI \cite{Schulze2011Two-tired}. Even though, CDHS proposed in \cite{BoraCDHS2016} and extended in \cite{ASCOM18} consider identical planetary parameters such as Mass, Radius, Escape Velocity and Surface Temperature; the contrasts are overwhelming compared to the similarities. The note aims to bring out the differences and investigates the contrasts between the two metrics, both from machine learning and modeling perspectives.

It is worth mentioning that once we know one observable -- the mass --  other planetary parameters used in the ESI computation (radius, density and escape velocity) can be calculated based on certain assumptions. For example, the small mass of Proxima~b suggests a rocky composition. However, since 1.27 EU is only a low limit on mass, it is still possible that its radius exceeds 1.5 -- 1.6 EU, which would make Proxima~b not rocky . Since Proxima~b mass is 1.27 EU, the radius is $R=M^{0.5} \equiv 1.12$ EU\footnote{http://phl.upr.edu/library/notes/standardmass-radiusrelationforexoplanets, Standard Mass-Radius Relation for Exoplanets, Abel Mendez, June 30, 2012.}. Accordingly, the escape velocity was calculated by $V_e=\sqrt{2GM/R}\equiv 1.065$ (EU), and the density by the usual $D=3M/4\pi R^3\equiv 0.904$ (EU) formula.

The manuscript, \cite{ASCOM18} consists of three related analyses: (i) computation and comparison of ESI and CDHS habitability scores for Proxima-b and the Trappist-1 system, (ii) some considerations on the computational methods for computing the CDHS score, and (iii) a machine learning exercise to estimate temperature-based habitability classes. The analysis is carefully conducted in each case, and the depth of the contribution to the literature helped unfold the differences and approaches to CDHS and ESI.\\

Several important characteristics were introduced to address the habitability question. \cite{Schulze2011Two-tired} first addressed this issue through two indices, the Planetary Habitability Index (PHI) and the Earth Similarity Index (ESI), where maximum, by definition, is set as 1 for the Earth, PHI=ESI=1.  

ESI represents a quantitative measure with which to assess the similarity of a planet with the Earth on the basis of mass, size and temperature. But ESI alone is insufficient to conclude about the habitability, as planets like Mars have ESI close to 0.8 but we cannot still categorize it as habitable. There is also a possibility that a planet with ESI value slightly less than 1 may harbor life in some form which is not there on Earth, i.e. unknown to us. PHI was quantitatively defined as a measure of the ability of a planet to develop and sustain life. However, evaluating PHI values for large number of planets is not an easy task. Irwin in 2014, introduced another parameter was  to account for the chemical composition of exoplanets and some biology-related features such as substrate, energy, geophysics, temperature and age of the planet --- the Biological Complexity Index (BCI). Here, we briefly describe the mathematical form of ESI. This will help us understand the difference in formulation of ESI and CDHS subsequently.
 
\paragraph{\bf Earth Similarity Index (ESI)}

ESI was designed to indicate how Earth-like an exoplanet might be \cite{Schulze2011Two-tired} and is an important factor to initially assess the habitability measure. Its value lies between 0 (no similarity) and 1, where 1 is the reference value, i.e. the ESI value of the Earth, and a general rule is that any planetary body with an ESI over 0.8 can be considered as Earth-like. It was proposed in the form 
\begin{equation}
ESI_{x}= \left( 1-\left | \frac{x-x_{0}}{x+x_{0}} \right | \right)^{w}\,,
\label{eq:ESIgeneral}
\end{equation}
where ESI$_x$ is the ESI value of a planet for $x$ property, and $x_{0}$ is the Earth's value for that property. The final ESI value of the planet is obtained by combining the geometric means of individual values, where $w$ is the weighting component through which the sensitivity of scale is adjusted. Four parameters: surface temperature $T_s$, density $D$, escape velocity $V_e$ and radius $R$, are used in ESI calculation. This index is split into interior $ESI_i$ (calculated from radius and density), and surface $ESI_s$ (calculated from escape velocity and surface temperature). Their geometric means are taken to represent the final $ESI$ of a planet. However, ESI in the form (\ref{eq:ESIgeneral}) was not introduced to define habitability, it only describes the similarity to the Earth in regard to some planetary parameters. For example, it is relatively high for the Moon. 

\textit{What's the paradox?}\\
The ESI and CDHS scores should be similar in the sense of being computed from essentially the same ingredients. However, similarity in numerical values may be accidental also unless  more informative picture could be provided with the additional exploration of the general relationship between ESI and CDHS scores. We explore relationships and establish the contrasts in the following section. It will also be established that ESI and CDHS are not related and there exists no causal relationship between the two.

CDHS is NOT derived from ESI. CDHS is NOT a classifier, neither is ESI !!!!!!!!!! Moreover, as new features and/or observables are added, there is no known estimate of the behavior of ESI, numerically speaking. As ESI is not based on maximizing a score, when we go on increasing the number of constituent parameters, the numerical value will go on diminishing. The reason for this is that each constituent term in the ESI is a number between 0 and 1 -- now as we add more terms to this, the value of the score tends to zero. This is not the case with our model, CDHS, as we have shown in the supplementary file, \cite{ASCOM18} that global optima is unaffected under additional input parameters (finitely mnay). However this throws a computational challenge of local oscillations which has been tackled by Stochastic Gradient Ascent in the same paper.\\
However, we do not intend to argue that the CD-HPF is a superior metric. However, we do argue that the foundations of the CD-HPF, in both mathematical and philosophical sense, are solidly grounded, and substantiated by analytical proofs.\\
\begin{enumerate}[a.]

\item The CDHS is \textit{not} derived from ESI. CDHS is not a \textit{classifier} and neither is ESI -- this is because we do not (yet) categorize planets based on the values of either ESI or CDHS. Even though we were categorical in emphasizing the contrast and reconciliation of two approaches (one being CDHS), we reiterate our baseline argument for the benefit of the readership. CDHS contributes to the Earth-Similarity concepts where the scores have been used to classify exoplanets based on their degree of similarity to Earth. However Earth-Similarity is not equivalent to exoplanetary habitability. Therefore, we adopted another approach where machine classification algorithms have been exploited to classify exoplanets in to three classes, non-habitable, mesoplanets and psychroplanets. While this classification was performed, CDHS was not used at all, rather discriminating features from the PHL-EC were used. This is fundamentally different from CDHS based Earth-Similarity approach where explicit scores were computed. Therefore, it was pertinent and remarkable that the outcome of these two fundamentally distinct exercises reconcile. We achieved that. This reconciliation approach is the first of its kind and fortifies CDHS, more than anything else. The manuscript, \cite{ASCOM18} captures this spirit. We maintain that this convergence between the two approaches is not accidental. We have been constantly watching the catalog, PHL-EC and scientific investigations in habitability of exoplanets. Please refer to the discussion section of the revised manuscript mentioned above for further details. We urge the readership to revisit the schematic flow elucidated in Fig. 1 of the manuscript. 

\item As an exercise, we tried to find the optimal point of the ESI in a same fashion as we found it for CD-HPF -- the finding was that a minima or a maxima cannot be guaranteed by the functional form of the ESI.

\item As ESI is not based on maximizing a score, when we go on increasing the number of constituent parameters, the numerical result of the ESI might go on decreasing: there is no guarantee of stability. The reason for this is that each constituent term in the ESI is a number between 0 and 1 -- now as we add more terms to this, the value of the score tends to zero. This is not the case with our model, CD-HPF, as we have shown in the supplementary file of \cite{ASCOM18} that global optima is unaffected under additional input parameters (finitely many). However this throws a computational challenge of local oscillations which has been tackled by Stochastic Gradient Ascent\cite{ASCOM18}.

\item It is easy to misconstrue CDHS with ESI as being one since the parameters of the CD-HPF are the same as that of the ESI, and the functional form in either metric is multiplicative in nature. However, that is not the case. We have laid a little emphasis on how our metric is similar to ESI in saying that the ESI is a special case of the CD-HPF. The crux of the philosophy behind the CD-HPF is essentially that of \textit{adaptive modeling}, i.e., the score generated is based on the best combination of the factors and not by static weights as followed by ESI.

\begin{itemize}
\item That, essentially, the ESI score gives non-dynamic weights to all the different planetary (with no trade-off between the weights) observables or calculated features considered, which in practice may not be the best approach, or at least, the only way of indicating habitability. It might be reasonable to say that for different exoplanets, the various planetary observables may weigh each other out to create a unique kind of favorable condition. For instance, in one planet, the mass may be optimal, but the temperature may be higher than the average of the Earth, but still within permissible limits; in another planet, the temperature may be similar to that of the Earth, but the mass may be lower. By discovering the best combination of the weights (or, as we call it, \textit{elasticities}) to maximize the resultant score, to the different planetary observables, we are creating a score which presents the best case scenario for the habitability of a planet. \\


\end{itemize}

\begin{table}[ht!]
\caption{\textbf{Differences between ESI and CDHS at a glance}}
\begin{center}
\begin{tabular}{ | m{0.5em} | m{4cm}| m{4cm} | } 
\hline
\textbf{S. No.}  & \textbf{ESI} & \textbf{CDHS}\\
\hline
1.   & Derived for four input parameters only. It is unclear how the ESI will behave if additional parameters such as eccentricity, flux, radial velocity, etc. are added. & On the contrary, the CDHS is solidly grounded in optimization theory as we have shown (in the supplementary file of \cite{ASCOM18}).\\
\hline
2. & The exponents of each term in the CDHS, $w_i/n$ is predetermined. & The exponents $\alpha$, $\beta$, $\gamma$ and $\delta$ are not predetermined; computing them is a part of the optimization problem.\\ 
\hline
3. & In all likelihood, the inclusion of additional input parameters will diminish the ESI since all input parameters are scaled between 0 and 1, and the weights are fixed. & This does not happen with the CDHS even if we include additional parameters.\\ 
\hline
\end{tabular}
\end{center}
\end{table}


\end{enumerate}

\textbf{Key Differences between ESI and CDHS:} In the case of the ESI, there is no evidence of a rigorous functional analysis. Schulze-Makuch et al. have not reported the existence of extrema for the ESI. We tried to find the maximum for the ESI in a manner similar to that of CD-HPF and we were unable to do so with the appropriate mathematical tools. This means that if we were to draw parallels between the CDHS and the ESI, the ESI would have no guarantee of having a maxima. The CD-HPF is based on an adaptive modeling, that is, when the CDHS is computed, the response is a maximum, which is based on the functional form of the CD-HPF. We reported the proof in order to substantiate the fact that a maximum does exist of the functional form that we have used. The highlights of the CD-HPF are:

\begin{enumerate}
\item The exponents (or, the \textit{elasticities}) of each observable in the function, $R$, $D$, $T_s$, $V_e$ are denoted using $\alpha$, $\beta$, $\gamma$, and $\delta$ respectively. The constraints on the permissible values of the elasticities are: -- 
$$
\alpha + \beta + \gamma + \delta \leq 1
$$
and
$$
\alpha, \beta, \gamma, \delta \geq 0
$$

Thus, the computation of the CDHS of a planet is essentially a \textit{constrained optimization problem}.

\item In \cite{ASCOM18}, we have also shown that the number of components in the CDHS can be countably infinite. Of course, the components being considered for each planet should be the same and the scoring system becomes different, but the possibility still exists that $n$ number of observables and/or input parameters can be accommodated into the function, while keeping the constraints on the elasticities intact. Moreover, we have proved a very important theorem that even if $n$ number of observables and/or input parameters make the functional form extremely complicated, a global optima is still guaranteed.
\end{enumerate}

\begin{figure}[h!]
\includegraphics[width=0.55\textwidth,height=0.35\textheight]{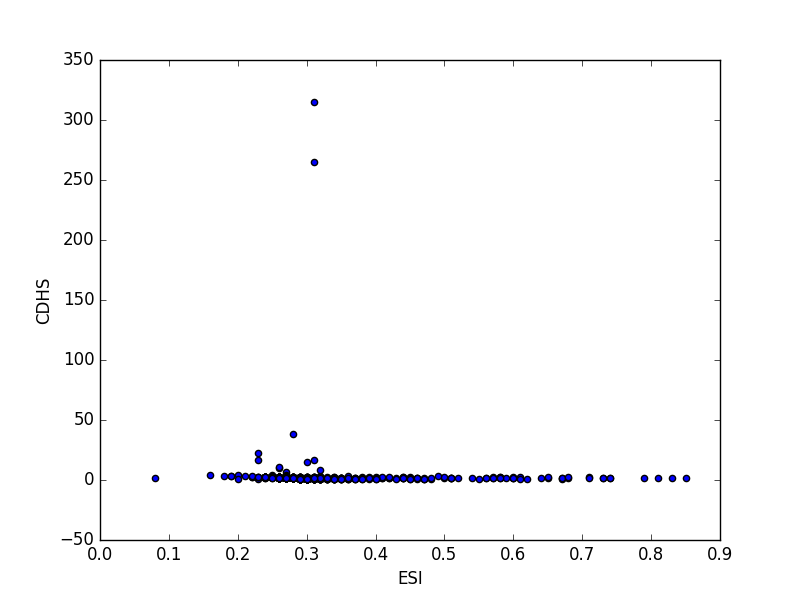}
\caption{ESI vs CDHS(CRS)}
\includegraphics[width=0.5\textwidth,height=0.35\textheight]{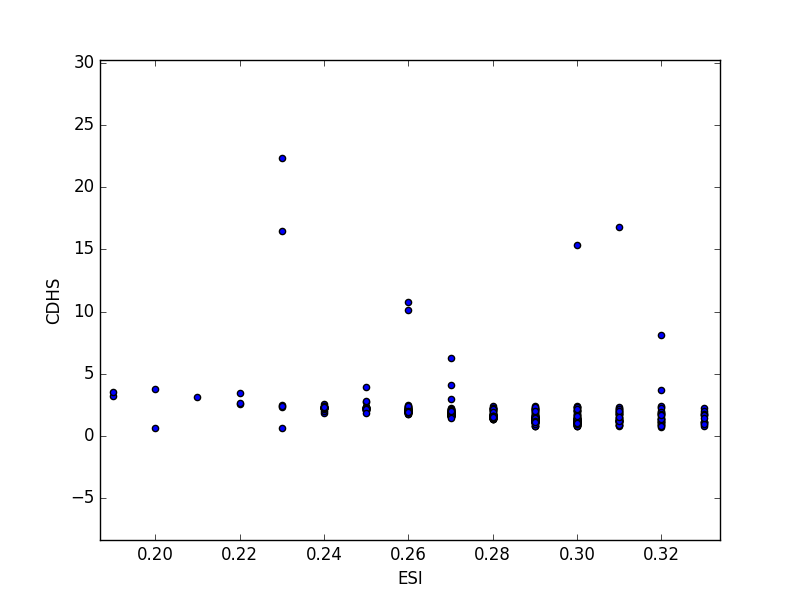}
\caption{ESI vs CDHS (CRS) -- a closeup}
      \label{esi-cdhs-crs-2}
\end{figure}

\begin{figure}[h!]
\includegraphics[width=0.55\textwidth,height=0.35\textheight]{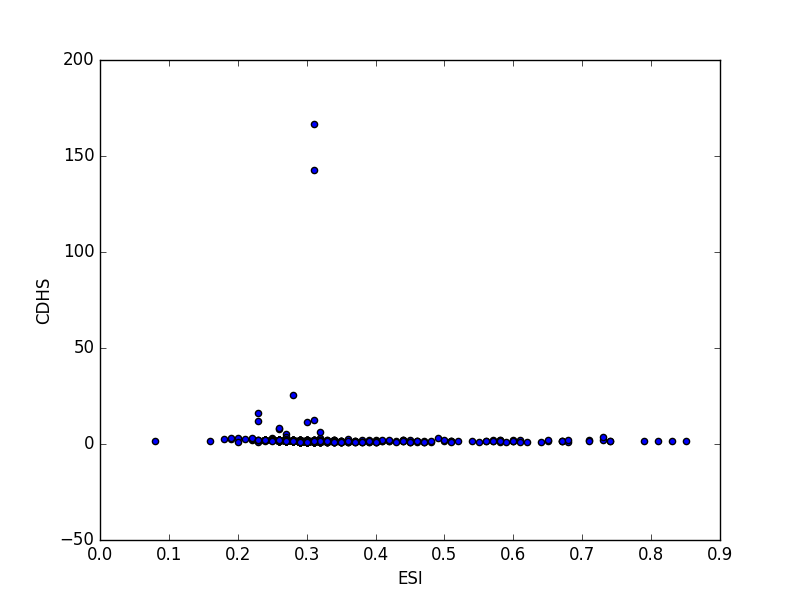}
\caption{ESI vs CDHS(DRS)}
\includegraphics[width=0.5\textwidth,height=0.35\textheight]{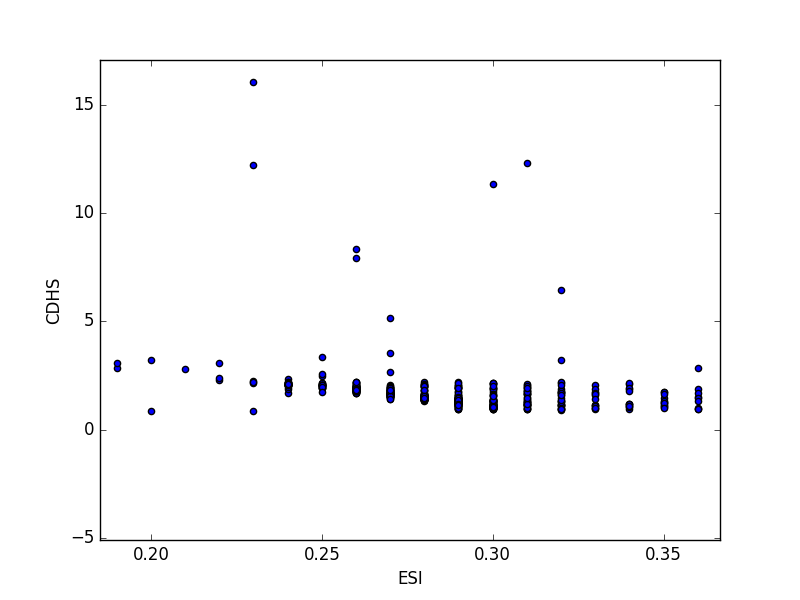}
\caption{ESI vs CDHS (CRS) -- a closeup}
      \label{esi-cdhs-drs-2}
\end{figure}

The reason why CDHS computation is a challenging problem (and ESI is not!) is
\begin{itemize}
\item Local oscillations about the optima \cite{Saha2016optimization}, \cite{GindeEconometric}, \cite{SFA} are difficult to mitigate even though we have shown there exists a theoretical guarantee for the same.
\item Extrema doesn't occur at the corner points and is therefore the location of such is difficult to predict.
\item We have emphasized enough on how the CD-HPF reconciles with the machine learning methods that we have used to automatically classify exoplanets. It is not easy two  to propose two fundamentally different approaches (one of which is CDHS) that lead to a similar conclusion about an exoplanet. While the CDHS provides a numerical indicator (in fact the existence of one global optima shouldn't be a concern at all,  rather a vantage point of the model that this eliminates the possibility of computing scores arbitrarily), the machine classification bolsters our proposition by telling us automatically which class of habitability an exoplanet belongs to. 
The performance of  machine classification is evaluated by class-wise accuracy. The accuracies achieved are remarkably high, and at the same time, we see that the values of the CDHS for the sample of potentially habitable exoplanets which we have considered are also close to 1. Therefore, the computational approaches map earth similarity to habitability. This is remarkable and non-trivial. This degree of computational difficulty is non-existent in ESI.

\item \textbf{NOTE:} One should not miss the crucial point of computation of CDHS which is only a part of the exercise. The greater challenge is the vindication of the CDHS metric in the classification of habitability of exoplanets by reconciliation of the modeling approach with the machine classification approach (where explicit scores were not used to classify habitability of exoplanets but the outcome validates the metric). The greatest strength of CDHS is its flexibility in functional form helping forge the unification paradigm.
\end{itemize}

To be specific, even under the constrained optimization problem, the search space is countably infinite making the problem of finding the global optima computationally intractable until some intervention is made. We have mitigated the problem by employing stochastic gradient ascent. If it's a simplex problem, we know the optima is at one of the corner points (theoretically) and therefore don't bother looking for it in the entire search space (please note even in that case, the computational complexity is hard to ignore and people develop different kinds of approximation algorithms to improve upon the complexity). The functional form in our case is non-linear, convex and is bounded by constraints (which is typically a geometric region, contour/curved) making the search space complex enough to find the optima. The only positive thing in our favor is the theoretical existence of global optima which ensures, once our algorithm finds the optima, it terminates! There is the other issue of handling oscillation around the optima, coupled with finding "the optima". 

It is, indeed the question of finding the optima efficiently that distinguishes CDHS from ESI. As an exploratory investigation, we rendered ESI a functional form, similar to CDHS with the express objective of finding elasticity (dynamic weights to each of the parameters) that may compute optimum Earth similarity of any exoplanet. What we found was such theoretical guarantee is non-existent for ESI and it may ensure global optima only in the case of IRS, which is infeasible (Please refer to \cite{BoraCDHS2016}). Therefore, our attempt to render theoretical credence to ESI fails as the new metric holds for IRS condition only (neither maxima nor minima). The new ESI input structure, despite a functional form very similar to CDHS is not able to reproduce the dynamic and flexible behavior of the Earth Similarity Score reported by CDHS. The details are documented in Appendix A.

\section{Discussion}

While each of these analyses is conducted independently, the results reconcile in a way that is sensible. The earth similarity indices, usually a solution that doesn't address the habitability classification problem has been reconciled with machine classification approach. In the process of doing so, we developed CDHS which is more representative (as opposed to ESI) of the mapping between these two completely different approaches.

A different perspective of the inference sought from the results of the ML algorithms lies in the values of the Cobb-Douglas Habitability Scores (CDHS). We see that the values of CDHS of Proxima-b and TRAPPIST-1 c, d, and e (on which the probability of life is deemed to be more probable), \cite{ASCOM18} are closer to the CDHS of Earth \footnote{CDHS  for MARS = 0.583 for all the variation of S.Temp; CDHS for MOON = 0.336, ESI of MARS=0.64, ESI of MOON=0.56} than those of the remaining planetary samples which we have specifically explored in our study.

The exposition of the efficacy of our methods are concurrently being made by astrophysicists. In a recent article in Astrobiology, it has been said that two planets in the TRAPPIST-1 system are likely to be habitable: \href{http://astrobiology.com/2018/01/two-trappist-1-system-planets-are-potentially-habitable.html}{http://astrobiology.com/2018/01/two-trappist-1-system-planets-are-potentially-habitable.html}. TRAPPIST-1 b and c are likely to have molten-core mantles and rocky surfaces, and hence, moderate surface temperatures and modest amounts of tidal heating. The expected favorable conditions are also reflected in the CDHS values of these planets as reported in the 2018 paper.

The computational aspect is also worth mentioning here so as to provide the reader a thorough understanding of the method we have explored, their advantages over existing methods, and notwithstanding, their limitations.

The essence of the CD-HPF, and consequently, that of the CDHS is indeed orthogonal to the essence of the ESI or BCI. We argue not in favor of the superiority of our metric, but for the new approach that is developed. We believe that there should actually be various metrics arising from different schools of thought so that the habitability of an exoplanet may be collectively determined from all these. Such a kind of adaptive modeling has not been used in the context of planetary habitability prior to the CD-HPF. While not in agreement with the structure of ESI, CDHS does complement ESI.

The rate at which exoplanets are being discovered is rapidly increasing. Given the current situation and how technology is rapidly improving in astronomy, we firmly believe that eventually, problems such as this in observational astronomy will require a thorough dealing of data science to be handled efficiently. As the number of confirmed exoplanets grow, it might be wise to use an indexing and classification method which can do the job automatically, with little need for human intervention. This is where both ESI and CDHS shall play useful roles, we believe.

\par
\textbf{{Appendix A: Constraint Conditions for elasticity; $\alpha$, $\beta$, $\gamma$ and $\delta$: ESI with dynamic input elasticity fails to be an optimizer.}}\\ 
In the function,
\begin{equation}
\begin{split}
Y= k.\left(1- \frac{R_e-R}{R_e+R}\right)^\alpha.\left(1- \frac{D_e-D}{D_e+D}\right)^\beta.\\
\left(1- \frac{T_e-T}{T_e+T}\right)^\gamma.\left(1- \frac{V_e-V}{V_e+V}\right)^\delta 
\end{split}
\end{equation}

where \(R_{e}\), \(D_{e}\),\(T_{e}\) and \(V_{e}\) are the radius, density, surface temperature and escape velocity of Earth and are constant terms. R,D.T and V are radius , density ,surface temperature and escape velocity for the planet under study.k is also a constant parameter
\\
Differentiating eq.1 above partially w.r.t. R
\begin{equation}
\frac{\partial Y}{\partial R}= \alpha.\left(\frac{2R}{R_e+R}\right)^{\alpha-1}.\frac{2R_e}{(R_e+R)^2}
\end{equation}
finding second derivative from eq.2

\begin{dmath}
\frac{\partial ^2 Y}{\partial R^2} =
\alpha.\left(\alpha-1\right).\left(\frac{2R}{R_e+R}\right)^{\alpha -2}.\left(\frac{2R_e}{(R_e+R)^2}\right).\left(\frac{2R_e}{(R_e+R)^2}\right)- \alpha.\left(\frac{2R}{R_e+R}\right)^{\alpha -1}.\left(\frac{4R_e(R_e+R)}{(R_e+R)^4}\right) \end{dmath} 
\begin{dmath}
 = \alpha.\left(\alpha-1\right).\left(\frac{2R}{R_e+R}\right)^{\alpha -2}.\frac{4R_e^2}{(R_e+R)^4}-\alpha.\left(\frac{2R}{R_e+R}\right)^{\alpha -1}.\left(\frac{4R_e^2+4R_e.R)}{(R_e+R)^4}\right) \end{dmath}
 \begin{dmath}
= \alpha.\left(\alpha-1\right).\left(\frac{2R}{R_e+R}\right)^{\alpha -2}.\frac{1}{(R_e+R)^4}.\left((\alpha-1).4R_e^2-\frac{2R}{R_e+R}.(4R_e^2 + 4RR_e)\right) \end{dmath}
\begin{dmath}
=\alpha.\left(\alpha-1\right).\left(\frac{2R}{R_e+R}\right)^{\alpha -2}.\frac{1}{(R_e+R)^4}.\left((\alpha-1).4R_e^2-8RR_e\right)\end{dmath}

hence final second order derivative is-
\begin{equation}
\frac{\partial ^2 Y}{\partial R^2}= \alpha.\left(\alpha-1\right).\left(\frac{2R}{R_e+R}\right)^{\alpha -2}.\frac{1}{(R_e+R)^4}.\left((\alpha-1).4R_e^2-8RR_e\right)
\end{equation}
for concavity eq.3 must be greater than 0
so, 
\begin{align}
\alpha.\left(\alpha-1\right).\left(\frac{2R}{R_e+R}\right)^{\alpha -2}.\frac{1}{(R_e+R)^4}.\left((\alpha-1).4R_e^2-8RR_e\right) > 0
\end{align}
on solving above inequality we will get ,
\begin{align}
\alpha -1 > 2 \frac{R}{R_e}
\end{align}
similarly it can be proved for other 3 elasticity constants which gives us constrained conditions
\begin{align}
\beta -1 > 2 \frac{D}{D_e}, & \\
\gamma -1 > 2 \frac{T}{T_e}, & \\
\delta -1 > 2 \frac{V}{V_e}
\end{align}
Summing up equations (10),(11,(12) and (13),
\begin{align}
\alpha +\beta +\gamma +\delta -4 > 2\left(\frac{R}{R_e}+\frac{D}{D_e}+\frac{T}{T_e}+\frac{V}{V_e}\right) \\
\Rightarrow \alpha +\beta +\gamma +\delta  > 2\left(\frac{R}{R_e}+\frac{D}{D_e}+\frac{T}{T_e}+\frac{V}{V_e}\right) + 4
\end{align}
Above Equation (10) shows that sum of the four elasticity constants cannot be less than or equal to 1 (in fact cannot be less than $1$). This is the case of IRS (increasing return to scale) in CDHPF function, which means that function is neither concave nor convex. The new metric holds for IRS condition only which doesn't ensure a global maxima implying lack of theoretical foundation for the ESI input structure.
\par
\textbf{{Appendix B: Optimization proof using Lagrangian Multiplier.}}\\ 
Here we provide the analytical proof of the claim made that this is the case of IRS (increasing return to scale) in CDHPF function, which means that function is neither concave nor convex. The new metric holds for IRS condition only, which doesn't ensure a global maxima.
The function Y is given as specified in eq.(2) in Appendix A, reproduced here for ease-
\begin{equation}
\begin{split}
Y= k.\left(1- \frac{R_e-R}{R_e+R}\right)^\alpha.\left(1- \frac{D_e-D}{D_e+D}\right)^\beta.\\
\left(1- \frac{T_e-T}{T_e+T}\right)^\gamma.\left(1- \frac{V_e-V}{V_e+V}\right)^\delta 
\end{split}
\end{equation}
It can be rewritten as-
\begin{equation}
\begin{split}
Y= k.\left(\frac{2R}{R_e+R}\right)^\alpha.\left(\frac{2D}{D_e+D}\right)^\beta.\\
\left(\frac{2T}{T_e+T}\right)^\gamma.\left(\frac{2V}{V_e+V}\right)^\delta 
\end{split}
\end{equation}
The Lagrangian function for the optimization problem is hence given as-

\begin{equation}
\begin{split}
\cal L&= Y-\lambda(w_1\left(\frac{2R}{R_e+R}\right)+w_2\left(\frac{2D}{D_e+D}\right)\\
&+w_3\left(\frac{2T}{T_e+T}\right)+
w_4\left(\frac{2V}{V_e+V}\right)-m) 
\end{split}
\end{equation}
\begin{align*}
&=k.\left(\frac{2R}{R_e+R}\right)^\alpha.\left(\frac{2D}{D_e+D}\right)^\beta.\\ \nonumber &
\left(\frac{2T}{T_e+T}\right)^\gamma.
\left(\frac{2V}{V_e+V}\right)^\delta -\lambda(w_1\left(\frac{2R}{R_e+R}\right)+ \\ & w_2\left(\frac{2D}{D_e+D}\right)+w_3\left(\frac{2T}{T_e+T}\right)+\\ \nonumber &
w_4\left(\frac{2V}{V_e+V}\right)-m) 
\end{align*}
The first order conditions are-
\begin{equation}
\begin{split}
\frac{\partial \cal L}{\partial R} =k\alpha \left(\frac{2R}{R_e+R}\right)^{\alpha-1}\left(\frac{2D}{D_e+D}\right)^\beta \\ 
\left(\frac{2T}{T_e+T}\right)^\gamma  \left(\frac{2V}{V_e+V}\right)^\delta-w_1\lambda =0\\
\end{split}
\end{equation}
\begin{equation}
\begin{split}
\frac{\partial \cal L}{\partial D}= k\alpha \left(\frac{2R}{R_e+R}\right)^{\alpha}\left(\frac{2D}{D_e+D}\right)^{\beta-1}\\ \left(\frac{2T}{T_e+T}\right)^\gamma \left(\frac{2V}{V_e+V}\right)^\delta-w_2\lambda =0\\
\end{split}
\end{equation}
\begin{equation}
\begin{split}
\frac{\partial \cal L}{\partial T}= k\alpha \left(\frac{2R}{R_e+R}\right)^{\alpha}\left(\frac{2D}{D_e+D}\right)^\beta\\ \left(\frac{2T}{T_e+T}\right)^{\gamma-1} \left(\frac{2V}{V_e+V}\right)^\delta-w_3\lambda =0\\
\end{split}
\end{equation}
\begin{equation}
\begin{split}
\frac{\partial \cal L}{\partial V}= k\alpha \left(\frac{2R}{R_e+R}\right)^{\alpha}\left(\frac{2D}{D_e+D}\right)^\beta \\
\left(\frac{2T}{T_e+T}\right)^\gamma \left(\frac{2V}{V_e+V}\right)^{\delta-1}-w_4\lambda =0\\
\end{split}
\end{equation}
\begin{equation}
\begin{split}
 \frac{\partial \cal L}{\partial \lambda}=-(w_1\left(\frac{2R}{R_e+R}\right)+w_2\left(\frac{2D}{D_e+D}\right)+\\
w_3\left(\frac{2T}{T_e+T}\right)+w_4\left(\frac{2V}{V_e+V}\right)-m)=0
\end{split}
\end{equation}
From Eq. (19) and (20),
\begin{equation}
\begin{split}
\frac{\alpha.\left(\frac{2R}{R_e+R}\right)^{\alpha-1.}.\left(\frac{2D}{D_e+D}\right)^\beta}{w_1}=\frac{\beta.\left(\frac{2R}{R_e+R}\right)^{\alpha}.\left(\frac{2D}{D_e+D}\right)^{\beta-1}}{w_2}
\end{split}
\end{equation}
or
\begin{equation}
\begin{split}
\frac{2D}{D_e+D}=\frac{2R}{R_e+R}.\frac{\beta}{\alpha}.\frac{w_1}{w_2}
\end{split}
\end{equation}
Similarly, we get
\begin{equation}
\begin{split}
\frac{2T}{T_e+T}=\frac{2R}{R_e+R}.\frac{\gamma}{\alpha}.\frac{w_3}{w_2}
\end{split}
\end{equation}
and
\begin{equation}
\begin{split}
\frac{2V}{V_e+D}=\frac{2R}{R_e+R}.\frac{\delta}{\alpha}.\frac{w_4}{w_2}
\end{split}
\end{equation}
Putting the values from eq. (25),(26) and (27) in eq.(19) and solving it further we get,
\begin{equation}
\begin{split}
\frac{2R}{R_e+R} &=\left(\lambda.k.\alpha^{1-(\beta+\gamma+\delta)}.\beta^\beta.\gamma^\gamma .\delta^\delta\right)^{\frac{1}{1-\left(\alpha+\beta +\gamma+\delta \right)}}\\
&\times\left({w_1}^{\beta+\gamma+\delta-1} {w_2}^{-\beta}
{w_3}^{-\gamma}{w_4}^{-\delta}\right)^{\frac{1}{1-\left(\alpha+\beta +\gamma+\delta \right)}}
\end{split}
\end{equation}
Similarly, we get following
\begin{equation}
\begin{split}
\frac{2D}{D_e+D} &=\left(\lambda.k.\beta^{1-(\alpha+\gamma+\delta)}.\alpha^\alpha.\gamma^\gamma .\delta^\delta\right)^{\frac{1}{1-\left(\alpha+\beta +\gamma+\delta \right)}}\\
&\times\left({w_2}^{\alpha+\gamma+\delta-1} {w_1}^{-\alpha}
{w_3}^{-\gamma}{w_4}^{-\delta}\right)^{\frac{1}{1-\left(\alpha+\beta +\gamma+\delta \right)}}
\end{split}
\end{equation}
\begin{equation}
\begin{split}
\frac{2T}{T_e+T} &=\left(\lambda.k.\gamma^{1-(\alpha+\beta+\delta)}.\alpha^\alpha.\beta^\beta .\delta^\delta\right)^{\frac{1}{1-\left(\alpha+\beta +\gamma+\delta \right)}}\\
&\times\left({w_3}^{\alpha+\beta+\delta-1} 
{w_1}^{-\alpha}
{w_2}^{-\beta}
{w_4}^{-\delta}\right)^{\frac{1}{1-\left(\alpha+\beta +\gamma+\delta \right)}}
\end{split}
\end{equation}
\begin{equation}
\begin{split}
\frac{2V}{V_e+V} &=\left(\lambda.k.\delta^{1-(\alpha+\beta+\gamma)}.\alpha^\alpha.\beta^\beta .\gamma^\gamma\right)^{\frac{1}{1-\left(\alpha+\beta +\gamma+\delta \right)}}\\
&\times\left({w_4}^{\alpha+\beta+\gamma-1} 
{w_1}^{-\alpha}
{w_2}^{-\beta}
{w_3}^{-\gamma}\right)^{\frac{1}{1-\left(\alpha+\beta +\gamma+\delta \right)}}
\end{split}
\end{equation}

Therefore, Lagrangian function is given as
\begin{equation}
\begin{split}
\cal L&=w_1\left(\frac{2R}{R_e+R}\right)+w_2\left(\frac{2D}{D_e+D}\right)\\
&+w_3\left(\frac{2T}{T_e+T}\right)+
w_4\left(\frac{2V}{V_e+V}\right)- \lambda \left(f(R,D,T,V)-y'\right)
\end{split}
\end{equation}
For above function , 1st order conditions can be given as-
\begin{equation}
\begin{split}
\frac{\partial \cal L}{\partial R}=w_1\left(\frac{2R_e}{(R_e+R)^2}\right)-\lambda  k\alpha\left(\frac{2R}{R_e+R}\right)^{\alpha-1}\\ 
\left(\frac{2D}{D_e+D}\right)^\beta \left(\frac{2T}{T_e+T}\right)^\gamma \left(\frac{2V}{V_e+V}\right)^\delta=0
\end{split}
\end{equation}
\begin{equation}
\begin{split}
\frac{\partial \cal L}{\partial D}=w_2\left(\frac{2D_e}{(D_e+D)^2}\right)-\lambda  k\beta\left(\frac{2R}{R_e+R}\right)^{\alpha}\\ 
\left(\frac{2D}{D_e+D}\right)^{\beta-1} \left(\frac{2T}{T_e+T}\right)^\gamma \left(\frac{2V}{V_e+V}\right)^\delta=0
\end{split}
\end{equation}
\begin{equation}
\begin{split}
\frac{\partial \cal L}{\partial T}=w_3\left(\frac{2T_e}{(T_e+T)^2}\right)-\lambda  k\gamma\left(\frac{2R}{R_e+R}\right)^{\alpha}\\ 
\left(\frac{2D}{D_e+D}\right)^{\beta} \left(\frac{2T}{T_e+T}\right)^{\gamma-1}\left(\frac{2V}{V_e+V}\right)^\delta=0
\end{split}
\end{equation}
\begin{equation}
\begin{split}
\frac{\partial \cal L}{\partial V}=w_4\left(\frac{2V_e}{(V_e+V)^2}\right)-\lambda  k\delta\left(\frac{2R}{R_e+R}\right)^{\alpha}\\ 
\left(\frac{2D}{D_e+D}\right)^{\beta} \left(\frac{2T}{T_e+T}\right)^\gamma \left(\frac{2V}{V_e+V}\right)^{\delta-1}=0
\end{split}
\end{equation}
\begin{equation}
\begin{split}
\frac{\partial \cal L}{\partial \lambda}=k \left(\frac{2R}{R_e+R}\right)^\alpha \left(\frac{2D}{D_e+D}\right)^\beta \left(\frac{2T}{T_e+T}\right)^\gamma \\ \left(\frac{2V}{V_e+V}\right)^\delta-y'=0
\end{split}
\end{equation}
Putting the values from eq. (25),(26) and (27) in eq.(37) we get,
\begin{equation}
\begin{split}
y'=k.\left(\frac{2R}{R_e+R}\right)^{\alpha}\left(\frac{\beta}{\alpha}\frac{w_1}{w_2}\left(\frac{2R}{R_e+R}\right)\right)^\beta \\ \left(\frac{\gamma}{\alpha}\frac{w_1}{w_3}\left(\frac{2R}{R_e+R}\right)\right)^\gamma\left(\frac{\delta}{\alpha}\frac{w_1}{w_4}\left(\frac{2R}{R_e+R}\right)\right)^\delta\nonumber \\
\end{split}
\end{equation}
\begin{equation}
\begin{split}
y'=k\left(\frac{2R}{R_e+R}\right)^{\alpha+\beta+\gamma+\delta}\alpha^{-\beta-\gamma-\delta}\beta^{\beta}\gamma^{\gamma}\delta^{\delta}\\ w_1^{\beta+\gamma+\delta}w_2^{-\beta}w_3^{-\gamma}w_4^{-\delta}\nonumber\\
\end{split}
\end{equation}
On solving for radius, we get
\begin{equation}
\begin{split}
\frac{2R}{R_e+R}=\left(k^{-1}.\alpha^{\beta+\gamma+\delta}\beta^{-\beta}\gamma^{-\gamma}\delta^{-\delta}\right)^{\frac{1}{\alpha+\beta+\gamma+\delta}}\\ 
\times \left(w_{1}^{-\beta-\gamma-\delta}.w_{2}^\beta.w_{3}^\gamma.w_{4}^\delta.y' \right)^{\frac{1}{\alpha+\beta+\gamma+\delta}}
\end{split}
\end{equation}
Similarly, we have following calculations for other 3 parameters-
\begin{equation}
\begin{split}
\frac{2D}{D_e+D}=\left(k^{-1}.\beta^{\alpha+\gamma+\delta}\alpha^{-\alpha}\gamma^{-\gamma}\delta^{-\delta}\right)^{\frac{1}{\alpha+\beta+\gamma+\delta}}\\ 
\times \left(w_{2}^{-\alpha-\gamma-\delta}.w_{1}^\alpha.w_{3}^\gamma.w_{4}^\delta.y' \right)^{\frac{1}{\alpha+\beta+\gamma+\delta}}
\end{split}
\end{equation}
\begin{equation}
\begin{split}
\frac{2T}{T_e+T}=\left(k^{-1}.\gamma^{\alpha+\beta+\delta}\alpha^{-\alpha}\beta^{-\beta}\delta^{-\delta}\right)^{\frac{1}{\alpha+\beta+\gamma+\delta}}\\ 
\times \left(w_{3}^{-\alpha-\beta-\delta}.w_{1}^\alpha.w_{2}^\beta.w_{4}^\delta.y' \right)^{\frac{1}{\alpha+\beta+\gamma+\delta}}
\end{split}
\end{equation}
\begin{equation}
\begin{split}
\frac{2V}{V_e+V}=\left(k^{-1}.\delta^{\alpha+\beta+\gamma}\alpha^{-\alpha}\beta^{-\beta}\gamma^{-\gamma}\right)^{\frac{1}{\alpha+\beta+\gamma+\delta}}\\ 
\times \left(w_{4}^{-\alpha-\beta-\gamma}.w_{1}^\alpha.w_{2}^\beta.w_{3}^\gamma.y' \right)^{\frac{1}{\alpha+\beta+\gamma+\delta}}
\end{split}
\end{equation}
Cost for getting y' in optimized way is say C
where 
\begin{equation}
\begin{split}
C= w_1\left(\frac{2R}{R_e+R}\right)+w_2\left(\frac{2D}{D_e+D}\right)\\
+w_3\left(\frac{2T}{T_e+T}\right)+
w_4\left(\frac{2V}{V_e+V}\right)
\end{split}
\end{equation}
So, analytically C can be written as-
\begin{equation}
\begin{split}
C= Q. \left(w_{1}^{\alpha}w_{2}^{\beta}w_{3}^{\gamma}w_{4}^{\delta}.\right)^{\frac{1}{\alpha+\beta+\gamma+\delta}}. y^ {'\frac{1}{\alpha+\beta+\gamma+\delta}}
\end{split}
\end{equation}
where Q is given as-
\begin{equation}
\begin{split}
Q=k^{\frac{-1}{\alpha+\beta+\gamma+\delta}}\left[\frac{\alpha^{\beta+\gamma+\delta}}{\beta^\beta+\gamma^\gamma+\delta^\delta}+\frac{\beta^{\alpha+\gamma+\delta}}{\alpha^\alpha+\gamma^\gamma+\delta^\delta}\right]^{\frac{1}{\alpha+\beta+\gamma+\delta}}+ \\ k^{\frac{-1}{\alpha+\beta+\gamma+\delta}}\left[\frac{\gamma^{\alpha+\beta+\delta}}{\alpha^\alpha+\beta^\beta+\delta^\delta}+\frac{\delta^{\alpha+\beta+\gamma}}{\alpha^\alpha+\beta^\beta+\gamma^\gamma}\right]^{\frac{1}{\alpha+\beta+\gamma+\delta}}
\end{split}
\end{equation}
with $C_{avg}$ as
\begin{equation}
\begin{split}
C_{avg}= \frac{C}{y'}= Q\left[w_1^{\alpha}w_2^{\beta}w_3^{\gamma}w_4^{\delta}\right]^{\frac{1}{\alpha+\beta+\gamma+\delta}}.y^{'\left(\frac{1}{\alpha+\beta+\gamma+\delta}\right)-1}
\end{split}
\end{equation}
Deriving conditions for optimization-
\begin{equation}
\begin{split}
\lambda\alpha k\left(\frac{2R}{R_e+R}\right)^{\alpha-1}\left(\frac{2D}{D_e+D}\right)^\beta \left(\frac{2T}{T_e+T}\right)^\gamma\\  \left(\frac{2V}{V_e+V}\right)^\delta =w_1
\end{split}
\end{equation}
\begin{equation}
\begin{split}
\lambda\beta k\left(\frac{2R}{R_e+R}\right)^{\alpha}\left(\frac{2D}{D_e+D}\right)^{\beta-1} \left(\frac{2T}{T_e+T}\right)^\gamma\\ \left(\frac{2V}{V_e+V}\right)^\delta =w_2
\end{split}
\end{equation}
\begin{equation}
\begin{split}
\lambda\gamma k\left(\frac{2R}{R_e+R}\right)^{\alpha}\left(\frac{2D}{D_e+D}\right)^{\beta} \left(\frac{2T}{T_e+T}\right)^{\gamma-1}\\ \left(\frac{2V}{V_e+V}\right)^\delta =w_3
\end{split}
\end{equation}
\begin{equation}
\begin{split}
\lambda\delta k\left(\frac{2R}{R_e+R}\right)^{\alpha}\left(\frac{2D}{D_e+D}\right)^{\beta} \left(\frac{2T}{T_e+T}\right)^{\gamma}\\ \left(\frac{2V}{V_e+V}\right)^{\delta-1} =w_4
\end{split}
\end{equation}
Multiplying eq (46),(47),(48) and (49) by 
$\left(\frac{2R}{R_e+R} \right)$,$\left(\frac{2D}{D_e+D}\right)$,$\left(\frac{2T}{T_e+T}\right)$,$\left(\frac{2V}{V_e+V}\right)$ respectively,
\begin{equation}
\begin{split}
\lambda\alpha k\left(\frac{2R}{R_e+R} \right)^{\alpha}\left(\frac{2D}{D_e+D}\right)^\beta \left(\frac{2T}{T_e+T}\right)^\gamma\\ \left(\frac{2V}{V_e+V}\right)^\delta =w_1\left(\frac{2R}{R_e+R} \right)  
\end{split}
\end{equation}
\begin{equation}
\begin{split}
\lambda\beta k\left(\frac{2R}{R_e+R} \right)^{\alpha}\left(\frac{2D}{D_e+D}\right)^\beta \left(\frac{2T}{T_e+T}\right)^\gamma\\ \left(\frac{2V}{V_e+V}\right)^\delta =w_2\left(\frac{2D}{D_e+D} \right)  
\end{split}
\end{equation}
\begin{equation}
\begin{split}
\lambda\gamma k\left(\frac{2T}{T_e+T}\right)^{\alpha}\left(\frac{2D}{D_e+D}\right)^\beta \left(\frac{2T}{T_e+T}\right)^\gamma\\ \left(\frac{2V}{V_e+V}\right)^\delta =w_3\left(\frac{2T}{T_e+T}\right)  
\end{split}
\end{equation}
\begin{equation}
\begin{split}
\lambda\gamma k\left(\frac{2V}{V_e+V}\right)^{\alpha}\left(\frac{2D}{D_e+D}\right)^\beta \left(\frac{2T}{T_e+T}\right)^\gamma\\ \left(\frac{2V}{V_e+V}\right)^\delta =w_4\left(\frac{2V}{V_e+V}\right)  
\end{split}
\end{equation}
From eq(17), the above equations can be expressed as-
\begin{equation}
\begin{split}
\lambda\alpha.Y=w_1\left(\frac{2R}{R_e+R} \right)
\end{split}
\end{equation}
\begin{equation}
\begin{split}
\lambda\beta.Y=w_2\left(\frac{2D}{D_e+D} \right)
\end{split}
\end{equation}
\begin{equation}
\begin{split}
\lambda\gamma.Y=w_3\left(\frac{2T}{T_e+T} \right)
\end{split}
\end{equation}\begin{equation}
\begin{split}
\lambda\delta.Y=w_4\left(\frac{2V}{V_e+V} \right)
\end{split}
\end{equation}
\begin{equation}
\begin{split}
\lambda\alpha.Y=w_1\left(\frac{2R}{R_e+R} \right)
\end{split}
\end{equation}
From above equations, it can be concluded
\begin{equation}
\frac{2D}{D_e+D}=\frac{\beta}{\alpha}\frac{w_1}{w_2}\frac{2R}{R_e+R}
\end{equation}
\begin{equation}
\frac{2T}{T_e+T}=\frac{\gamma}{\alpha}\frac{w_1}{w_3}\frac{2R}{R_e+R}
\end{equation}
\begin{equation}
\frac{2V}{V_e+V}=\frac{\delta}{\alpha}\frac{w_1}{w_4}\frac{2R}{R_e+R}
\end{equation}
Putting the values from eq.(59),(60),(61) into eq.(46),(47),(48) and (49) and solving them we get following-
\begin{equation}
\begin{split}
\frac{2R}{R_e+R}=\left(\lambda.k\alpha^{1-\left(\beta+\gamma+\delta\right)}\beta^{\beta}\gamma^{\gamma}\delta^{\delta}\right)^{\frac{1}{1-\left(\alpha+\beta+\gamma+\delta\right)}}\\ \times \left(w_1^{\beta+\gamma +\delta-1}w_2^{-\beta}w_3^{-\gamma}w_4^{-\delta}\right)^{\frac{1}{1-\left(\alpha+\beta+\gamma+\delta\right)}}
\end{split}
\end{equation}
\begin{equation}
\begin{split}
\frac{2D}{D_e+D}=\left(\lambda.k\alpha^{\alpha}\beta^{1-\left(\alpha+\gamma+\delta\right)}\gamma^{\gamma}\delta^{\delta}\right)^{\frac{1}{1-\left(\alpha+\beta+\gamma+\delta\right)}}\\ \times \left(w_1^{-\alpha}w_2^{\alpha+\gamma +\delta-1}w_3^{-\gamma}w_4^{-\delta}\right)^{\frac{1}{1-\left(\alpha+\beta+\gamma+\delta\right)}}
\end{split}
\end{equation}
\begin{equation}
\begin{split}
\frac{2T}{T_e+T}=\left(\lambda k\alpha^{\alpha}\beta^{\beta}\gamma^{1-\left(\alpha+\beta+\delta\right)}\delta^{\delta}\right)^{\frac{1}{1-\left(\alpha+\beta+\gamma+\delta\right)}}\\ \times \left(w_1^{-\alpha}w_2^{-\beta}w_3^{\alpha+\beta +\delta-1}w_4^{-\delta}\right)^{\frac{1}{1-\left(\alpha+\beta+\gamma+\delta\right)}}
\end{split}
\end{equation}
\begin{equation}
\begin{split}
\frac{2V}{V_e+V}=\left(\lambda k\alpha^{\alpha}\beta^{\beta}\gamma^{\gamma}\delta^{1-\left(\alpha+\beta+\gamma\right)}\right)^{\frac{1}{1-\left(\alpha+\beta+\gamma+\delta\right)}} \\ \times \left(w_1^{-\alpha}w_2^{-\beta}w_3^{-\gamma}w_4^{\alpha+\beta +\gamma-1}\right)^{\frac{1}{1-\left(\alpha+\beta+\gamma+\delta\right)}} 
\end{split}
\end{equation}
These above obtained expressions are to be maximized. Substituting these values in function Y (given as),
\begin{equation}
\begin{split}
Y= \left(1- \frac{R_e-R}{R_e+R}\right)^\alpha.\left(1- \frac{D_e-D}{D_e+D}\right)^\beta.\\
\left(1- \frac{T_e-T}{T_e+T}\right)^\gamma.\left(1- \frac{V_e-V}{V_e+V}\right)^\delta 
\end{split}
\end{equation}
we obtain,
\begin{equation}
\begin{split}
Y=\left(kp^{\alpha+\beta+\gamma+\delta}\alpha^{\alpha}\beta^{\beta}\gamma^{\gamma}\delta^{\delta}\right)^{\frac{1}{1-\left(\alpha+\beta+\gamma+\delta\right)}}\\ \times \left(w_1^{-\alpha}w_2^{-\beta}w_3^{-\gamma}w_4^{-\delta}\right)^{\frac{1}{1-\left(\alpha+\beta+\gamma+\delta\right)}}
\end{split}
\end{equation}
If $\alpha+\beta+\gamma+\delta$ is 1, the model is undefined, which is the case of constant return to scale (CRS).In this case the sum of these elasticities must be greater than 1 for Y to be maximized (IRS) case. This is infeasible from earlier observations implying the CDHS with ESI input structure doesn't guarantee a global optima, certainly a drawback that CDHS doesn't have. 

\end{document}